\input harvmac
 

\def\coeff#1#2{\relax{\textstyle {#1 \over #2}}\displaystyle}
\def\ccoeff#1#2{{#1\over #2}}

\def\eql{~=~}
\def\seql{\! = \!}

\def\ga{\gamma}

\def\cF{{\cal F}} 
\def\cH{{\cal H}} \def\cI{{\cal I}}
 
\def\cL{{\cal L}} 
\def\cN{{\cal N}} \def\cO{{\cal O}}
\def\cP{{\cal P}} 
\def\cR{{\cal R}}

\def\ph{\phi}
\def\om{\omega}

\def\bfone{\relax{\rm 1\kern-.35em 1}}
\def\IC{\relax\,\hbox{$\inbar\kern-.3em{\rm C}$}}
\def\ID{\relax{\rm I\kern-.18em D}}
\def\IF{\relax{\rm I\kern-.18em F}}
\def\IH{\relax{\rm I\kern-.18em H}}
\def\II{\relax{\rm I\kern-.17em I}}
\def\IN{\relax{\rm I\kern-.18em N}}
\def\IP{\relax{\rm I\kern-.18em P}}
\def\IQ{\relax\,\hbox{$\inbar\kern-.3em{\rm Q}$}}
\def\us#1{\underline{#1}}
\def\IR{\relax{\rm I\kern-.18em R}}
\font\cmss=cmss10 \font\cmsss=cmss10 at 7pt
\def\ZZ{\relax\ifmmode\mathchoice
{\hbox{\cmss Z\kern-.4em Z}}{\hbox{\cmss Z\kern-.4em Z}}
{\lower.9pt\hbox{\cmsss Z\kern-.4em Z}}
{\lower1.2pt\hbox{\cmsss Z\kern-.4em Z}}\else{\cmss Z\kern-.4em
Z}\fi}
\def\th{\theta}
\def\ep{\epsilon}
\def\si{\sigma}

\def\rh{\rho}

\def\be{\beta}

\def\cI{{\cal I}}
\def\ga{\gamma}

\def\th{\theta}
\def\rh{\rho}
\def\be{\beta}
\def\cO{{\cal O}}

\def\ph{\phi}

%
\def\nihil#1{{``#1''}}

\def\nup#1({Nucl.\ Phys.\ $\us {B#1}$\ (}
\def\plt#1({Phys.\ Lett.\ $\us  {#1B}$\ (}
\def\cmp#1({Comm.\ Math.\ Phys.\ $\us  {#1}$\ (}
\def\prp#1({Phys.\ Rep.\ $\us  {#1}$\ (}
\def\prl#1({Phys.\ Rev.\ Lett.\ $\us  {#1}$\ (}
\def\prv#1({Phys.\ Rev.\ $\us  {#1}$\ (}
\def\mpl#1({Mod.\ Phys.\ Let.\ $\us  {A#1}$\ (}
\def\ijmp#1({Int.\ J.\ Mod.\ Phys.\ $\us{A#1}$\ (}
\def\jag#1({Jour.\ Alg.\ Geom.\ $\us {#1}$\ (}

\def\cI{1}

\lref\PilchEJ{
K.~Pilch and N.~P.~Warner,
``A new supersymmetric compactification of chiral IIB supergravity,''
Phys.\ Lett.\ B {\bf 487} (2000) 22,
hep-th/0002192.
}

\lref\PilchFU{
K.~Pilch and N.~P.~Warner,
``N = 1 supersymmetric renormalization group flows from IIB supergravity,''
Adv.\ Theor.\ Math.\ Phys.\  {\bf 4} (2002) 627,
hep-th/0006066.
}

\lref\KhavaevFB{
A.~Khavaev, K.~Pilch and N.~P.~Warner,
``New vacua of gauged N = 8 supergravity in five dimensions,''
Phys.\ Lett.\ B {\bf 487} (2000) 14,
hep-th/9812035.
}

\lref\CorradoNV{
R.~Corrado, K.~Pilch and N.~P.~Warner,
``An N = 2 supersymmetric membrane flow,''
Nucl.\ Phys.\ B {\bf 629} (2002) 74,
hep-th/0107220.
}

\lref\CorradoBZ{
R.~Corrado and N.~Halmagyi,
``N=1 Field Theories and Fluxes in IIB String Theory,''
hep-th/0401141.
}

\lref\CorradoWX{
R.~Corrado, M.~G\"unaydin, N.~P.~Warner and M.~Zagermann,
``Orbifolds and flows from gauged supergravity,''
Phys.\ Rev.\ D {\bf 65} (2002) 125024, hep-th/0203057.
}

\lref\KlebanovHH{
I.~R.~Klebanov and E.~Witten,
``Superconformal field theory on threebranes at a Calabi-Yau  singularity,''
Nucl.\ Phys.\ B {\bf 536} (1998) 199,
hep-th/9807080.
}

\lref\KlebanovHB{
I.~R.~Klebanov and M.~J.~Strassler,
``Supergravity and a confining gauge theory: Duality cascades and
$\chi$SB-resolution of naked singularities,''
JHEP {\bf 0008} (2000) 052, hep-th/0007191.
}

\lref\PilchJG{
K.~Pilch and N.~P.~Warner,
``Generalizing the N = 2 supersymmetric RG flow solution of IIB
supergravity,''
Nucl.\ Phys.\ B {\bf 675} (2003) 99,
hep-th/0306098.
}

\lref\FreedmanGP{
D.~Z.~Freedman, S.~S.~Gubser, K.~Pilch and N.~P.~Warner,
``Renormalization group flows from holography supersymmetry and a
c-theorem,''
Adv.\ Theor.\ Math.\ Phys.\  {\bf 3} (1999) 363,
hep-th/9904017.
}

\lref\PapadopoulosJK{
G.~Papadopoulos and D.~Tsimpis,
``The holonomy of IIB supercovariant connection,''
Class.\ Quant.\ Grav.\  {\bf 20} (2003) L253,
hep-th/0307127.
}

\lref\SchwarzQR{
J.~H.~Schwarz,
\nihil{Covariant field equations of chiral N=2 D=10 supergravity,}
Nucl.\ Phys.\ B {\bf 226} (1983) 269.
}

\lref\GauntlettSC{
J.~P.~Gauntlett, D.~Martelli, S.~Pakis and D.~Waldram,
``G-structures and wrapped NS5-branes,''
hep-th/0205050.
}

\lref\GauntlettNW{
J.~P.~Gauntlett, J.~B.~Gutowski, C.~M.~Hull, S.~Pakis and H.~S.~Reall,
``All supersymmetric solutions of minimal supergravity in five  dimensions,''
hep-th/0209114.
}

\lref\GauntlettFZ{
J.~P.~Gauntlett and S.~Pakis,
``The geometry of D = 11 Killing spinors,''
hep-th/0212008.
}

\lref\GauntlettWB{
J.~P.~Gauntlett, J.~B.~Gutowski and S.~Pakis,
``The geometry of D = 11 null Killing spinors,''
JHEP {\bf 0312} (2003) 049,
hep-th/0311112.
}

\lref\GauntlettDI{
J.~P.~Gauntlett,
``Branes, calibrations and supergravity,''
hep-th/0305074.
}

\lref\GowdigereJF{
C.~N.~Gowdigere, D.~Nemeschansky and N.~P.~Warner,
``Supersymmetric solutions with fluxes from algebraic Killing spinors,''
hep-th/0306097.
}

\lref\NemWar{
D.~Nemeschansky and N.~P.~Warner, ``A family of M-theory flows with four 
supersymmetries,'' USC-04/01, {\it to appear}.
}

\lref\LeighEP{
R.~G.~Leigh and M.~J.~Strassler,
``Exactly marginal operators and duality in four-dimensional N=1
supersymmetric gauge theory,''
Nucl.\ Phys.\ B {\bf 447} (1995) 95, 
hep-th/9503121.
}

\lref\TownsendCI{
P.~K.~Townsend,
``Killing spinors, supersymmetries and rotating intersecting branes,''
hep-th/9901102.
}

\lref\FigueroaVA{
J.~M.~Figueroa-O'Farrill,
``On the supersymmetries of anti de Sitter vacua,''
Class.\ Quant.\ Grav.\  {\bf 16} (1999) 2043,
hep-th/9902066.
}

\lref\KhavaevYG{
A.~Khavaev and N.~P.~Warner,
``An N = 1 supersymmetric Coulomb flow in IIB supergravity,''
Phys.\ Lett.\ B {\bf 522} (2001) 181,
hep-th/0106032.
}

\lref\JohnsonIC{
C.~V.~Johnson, K.~J.~Lovis and D.~C.~Page,
``Probing some N = 1 AdS/CFT RG flows,''
JHEP {\bf 0105} (2001) 036,
hep-th/0011166.
}

\lref\JohnsonZE{
C.~V.~Johnson, K.~J.~Lovis and D.~C.~Page,
``The K\"ahler structure of supersymmetric holographic RG flows,''
JHEP {\bf 0110} (2001) 014,
hep-th/0107261.
}

\lref\FreySD{
A.~R.~Frey and M.~Grana,
``Type IIB solutions with interpolating supersymmetries,''
Phys.\ Rev.\ D {\bf 68} (2003) 106002,
hep-th/0307142.
}

\lref\FigueroaTX{
J.~M.~Figueroa-O'Farrill,
``Breaking the M-waves,''
Class.\ Quant.\ Grav.\  {\bf 17} (2000) 2925,
hep-th/9904124.
}

\lref\MartelliKI{
D.~Martelli and J.~Sparks,
``G-structures, fluxes and calibrations in M-theory,''
Phys.\ Rev.\ D {\bf 68}  (2003) 085014,
hep-th/0306225.
}

\lref\NunezCF{
C.~Nunez, A.~Paredes and A.~V.~Ramallo,
``Flavoring the gravity dual of N = 1 Yang-Mills with probes,''
JHEP {\bf 0312} (2003) 024, 
 hep-th/0311201.
}
%

\lref\KhavaevGB{
A.~Khavaev and N.~P.~Warner,
 ``A class of N = 1 supersymmetric RG flows from 
 five-dimensional N = 8 supergravity,''
Phys.\ Lett.\ B {\bf 495}  (2000) 215,
 hep-th/0009159.
}

\lref\GranaXN{
M.~Grana and J.~Polchinski,
``Gauge / gravity duals with holomorphic dilaton,''
Phys.\ Rev.\ D {\bf 65}  (2002) 126005
 hep-th/0106014.
}

\lref\MyersPS{
R.~C.~Myers,
``Dielectric-branes,''
JHEP {\bf 9912} (1999) 022,
hep-th/9910053.
}

\lref\KarchPV{
A.~Karch, D.~Lust and A.~Miemiec,
``New N = 1 superconformal field theories and their supergravity
Phys.\ Lett.\ B {\bf 454}  (1999) 265,
 hep-th/9901041.
}

%

\Title{
\vbox{
\hbox{UB-ECM-PF-04/04}
\hbox{USC-04/02}
\hbox{\tt hep-th/0403005}
}}
{\vbox{\vskip -1.0cm
\centerline{{\hbox{$\cN\seql 1$ supersymmetric solutions of IIB supergravity}}}
\vskip 8 pt
\centerline{\hbox{ from Killing spinors}}}}
\vskip -.3cm
\centerline{Krzysztof Pilch${}^{(1),(2)}$%
\footnote{${}^\dagger$}{On sabbatical leave.}
and
Nicholas P.\ Warner${}^{(2)}$}
\bigskip
\centerline{{${}^{(1)}$\it Departament ECM, Facultat F\'\i sica }}
\centerline{{\it Universitat de Barcelona}}
\centerline{{\it Diagonal 647}}
\centerline{{\it 08028 Barcelona, Spain}}
\medskip
\centerline{{${}^{(2)}$\it Department of Physics and Astronomy}}
\centerline{{\it University of Southern California}}
\centerline{{\it Los Angeles, CA 90089-0484, USA}}

\bigskip \bigskip 

We present a new class of ``dielectric'' $\cN\seql 1$
supersymmetric solutions of IIB supergravity.  This class
contains not just  the ten-dimensional 
lift of the Leigh-Strassler renormalization group flow, but also
the Coulomb branch deformation of this flow in which the branes are
allowed to spread in a radially symmetric manner, preserving the
$SU(2)$ global symmetry.  We use the ``algebraic Killing spinor''
technique, illustrating how it can be adapted to $\cN=1$ supersymmetric
flows.

\vskip .3in
\Date{\sl {February, 2004}}
 
\vfill\eject

\newsec{Introduction}

Understanding and characterizing supersymmetric backgrounds with
fluxes is very important in string theory, particularly those
backgrounds that lead to $\cN\seql 1$ supersymmetry in four
dimensions.  In this paper we construct explicitly a new family of such
backgrounds in type IIB supergravity thereby extending the approach
in \refs{\PilchJG,\GowdigereJF} to backgrounds with less supersymmetry.

As in \refs{\PilchJG,\GowdigereJF}  the starting point  in our construction is a specific
Ansatz for the form of the Killing spinors, $\epsilon^{(i)}$, and the metric.  This
Ansatz is   motivated by the symmetry and general structure of the 
holographically dual  supersymmetric field theory in four dimensions. 
We then show that the  Killing spinor equations for unbroken supersymmetries 
together with the Bianchi  identities for  the fluxes completely determine the 
background in terms of a single 
``master function," $\Psi(u,v)$, satisfying a non-linear second order partial 
differential equation:
\eqn\themastereqn{
u^3 {\partial\over\partial u}\left({1\over u^3}\,
{\partial\over\partial u}\,\Psi\right)+
{1\over 2 v}\,
 {\partial\over\partial v}\left({1\over v^3}\,
 {\partial\over\partial v}\, e^{2\Psi}\right)\eql 0\,.
}
The fact that we find that the complete solution is characterized by
a single PDE should not be surprising:  We are seeking a solution that is dual 
to a field theory with a Coulomb branch and so one should expect to find some 
generalization of the ``harmonic rule'' that characterizes pure Coulomb branch flows. 
Our flow is massive and required to have an $SU(2)$ global symmetry:  Hence
the non-linearity and the reduction in the number of variables.

Our approach  is closely related to the recent work on the classification of 
supersymmetric backgrounds  in terms of $G$-structures (for a review see  \GauntlettDI). 
Since the analysis of those structures is still quite complicated, most of 
the work thus far has focussed  on   M-theory 
\refs{\GauntlettFZ,\GauntlettWB}  and on theories in lower dimensions and/or with fewer supersymmetries (see, for example, \ \refs{\GauntlettSC,\GauntlettNW}). 
In particular,  we are not aware of any systematic  study of $G$-structures in the IIB theory.

The basic objects in the $G$-structure approach 
are the differential $p$-forms
\eqn\sandwiches{\Omega^{ij}_{M_1M_2\ldots M_p}
 ~\equiv~ \bar \epsilon^{(i)} \gamma_{M_1M_2\ldots M_p}\epsilon^{(j)}\quad
 \hbox{and} \quad  C^{ij}_{M_1M_2\ldots M_p}
 ~\equiv~ \epsilon^{(i)}{}^TC \gamma_{M_1M_2\ldots M_p}\epsilon^{(j)}\,,}
constructed as bilinears in the Killing spinors.  Note that $\Omega$ involves
Dirac conjugation whereas $C$ involves the Majorana conjugate, which means 
that the former has
$\cR$-charge $0$ and the latter has $\cR$-charge $+1$. 
In particular, the one forms, $\Omega_M^{(ij)}$, 
give rise to  Killing vectors of the background \refs{\GauntlettFZ,\PilchJG}. 
This observation will prove crucial for the integration of the Killing spinor equations.
The forms \sandwiches\  satisfy a system of first order differential equations together 
with non-linear algebraic relations that follow from Fierz identities. 
Geometrically they encode the reduction of the holonomy of the supercovariant 
connection on the spinor bundle. It has been shown 
recently \PapadopoulosJK\  that 
this holonomy group for a type IIB background with Killing spinors
$\epsilon^{(i)}$, $i\eql 1,\ldots,\nu$, where $\nu$ is the number of unbroken 
supersymmetries between 0 and 32, is a subgroup of the semi-direct product 
$SL(32-\nu,R)\times (\oplus^\nu R^{32-\nu})$, in particular, 
 $SL(32,R)$ is the largest holonomy of a generic background.

Our emphasis here is on explicit construction of supersymmetric backgrounds 
with fluxes and we find it simpler to start directly with the invariant 
spinors of the $G$-structure.   More precisely, we make a very general
Ansatz for the metric  and to some extent for the fluxes, and then write some projection 
conditions that define the supersymmetries.  The supersymmetry variations then become 
(over-determined) algebraic equations for the background tensor-gauge fields, 
and the equations of motion then emerge from both the over-determined algebraic 
equations and the Bianchi identities for the field strengths.

In this paper we will find a family of solutions of IIB supergravity
with four supersymmetries that are also holographic duals of $\cN=1$
supersymmetric field theories in four dimensions.  In particular, we
find a family of solutions that generalizes the flows of
\refs{\PilchEJ,\PilchFU,\KhavaevYG}, which contain the holographic dual of 
a ``Leigh-Strassler'' (LS) renormalization
group flow \refs{\LeighEP,\FreedmanGP}.  We obtain a family of flows 
because the solution is
completely determined by a single function of two variables that is
required to satisfy \themastereqn.  
 Our restriction of the number of variables is
needed to make the problem manageable, but we expect that there should
be generalizations to more variables.  Our approach closely parallels
that of \NemWar\   in which families of solutions with four supersymmetries
are found in $M$-theory.  One of our purposes here is to show that the ideas of 
\refs{\PilchJG,\GowdigereJF} can be adapted, in a very simple manner, 
to address problems with less supersymmetry.  

Our approach is similar to others involving prescriptions for the spinors
that make up the supersymmetry.  The natural first step is to use the
Poincar\'e invariance on the brane to break the supersymmetry into
a ``$4+6$'' split \GranaXN:
\eqn\helsplit{ \epsilon ~=~ \zeta \otimes \chi^{(1)} ~+~ 
\zeta^*  \otimes {\chi^{(2)}}{}^* \,,}
where $\Gamma^{(4)} \zeta = +\zeta$ and $\Gamma^{(6)} \chi^{(i)}  = -\chi^{(i)}$
denote the helicity components in $4$ and $6$ dimensions respectively.
The issue of supersymmetry then hinges upon how $\chi^{(1)}$ is related
to $\chi^{(2)}$ \refs{\GranaXN\FreySD\MartelliKI\FigueroaTX
{--}\NunezCF}.  Interesting classes of solutions arise from relatively
simple relationships, such as  $\chi^{(2)} = 0$ (type B) or $\chi^{(2)} = 
e^{i \psi} \chi^{(1)}$ (types A and C).  However, the type of solution that we
wish to obtain, and that arises naturally in physically important massive  
flows of holographic gauge theories, do not fit into such simple schemes:
The relationship between  $\chi^{(1)}$ and $\chi^{(2)}$ is significantly
more complicated.

Fortunately, the underlying physics provides us with an invaluable
guide to solving the problem.  The massive flows we seek involve
fluxes for the $3$-form field strengths, and so one should expect some
dielectric polarization of the $D3$-branes into $D5$-branes through the
Myers effect \MyersPS.   One of the surprising results of  \PilchJG\ was that there
was a  concomitant ``dielectric'' deformation of the canonical supersymmetry projector.
That is, the presence of tensor gauge fields caused the supersymmetry
projector transverse to the original branes to be rotated so as to
receive a component in the internal directions:  The product $\gamma^{1234}$
was rotated into a term of the form  $\gamma^{1234AB}$, for some choice of $A$ and $B$.
We believe that this should be interpreted as polarizing some of the $D3$-branes 
into a mixture of $D5$ and $N\!S5$ branes.   We will find a very similar structure here 
for the flows with four supersymmetries, showing that such
dielectric deformations are an essential part of holographic RG flows.  

More generally,  we believe that it will be important to consider
the  dielectric deformation of the canonical  supersymmetry projector in
broader classes of string compactification, including backgrounds based
on compact manifolds.  In the context of holographic RG flows  
we  expect that our methods can be used to study other families of
supergravity solutions, in particular
(i) to find the explicit solution, and
generalizations, of the Klebanov-Witten flow \KlebanovHH, (ii) to test  the
conjecture duality of \refs{\CorradoWX, \CorradoBZ} that relates the 
flows that we obtained in
\refs{\PilchEJ,\PilchFU} to those of Klebanov and Witten \KlebanovHH, (iii)
to generalize the duality cascade of Klebanov and Strassler \KlebanovHB\ to
other UV fixed point theories.

In section 2 we describe our general Ansatz for the metric, tensor
gauge fields and supersymmetries.  While we can motivate our choice of
Ansatz rather generally, we actually arrived at it by a
detailed study of the supersymmetry in the solutions of
\refs{\PilchEJ,\PilchFU}.  To simplify the presentation, we describe
the particular solutions in section 4.  Section 3 contains the solution
to the general Ansatz, and section 5 contains some final remarks.
Throughout this paper we use the conventions of \SchwarzQR.

\newsec{The supersymmetry Ansatz}

\subsec{The underlying holographic field theory}

To understand and motivate the supergravity calculation it is useful
to recall some of the details of the holographic flow in field theory.
We consider a relevant deformation of ${\cal N}=4$ super-Yang-Mills
theory by a mass term for one of the three ${\cal N}=1$ adjoint chiral
superfields.  That is, we consider a superpotential of the
form
\eqn\gaugesuperpot{
 W ={\rm Tr} \,\big( \Phi_3
[\Phi_1,\Phi_2] \big) ~+~ \coeff{1}{2} m\, {\rm Tr}\big( \Phi_3^2\big)
\ .}
The first term is the superpotential inherited from the
original ${\cal N}=4$
supersymmetric gauge theory, while the second
term breaks conformal invariance, reduces the supersymmetry from
${\cal N}=4$ to ${\cal N}=1$, and drives the theory to a new,
non-trivial ${\cal N}=1$ superconformal fixed point in the infra-red
\LeighEP.
 
This infra-red fixed fixed point has a four-dimensional Coulomb branch
that may be described in terms of the vevs of the operators $\Phi_1$
and $\Phi_2$.  A two parameter family of flows on this Coulomb branch
were studied in \refs{\KhavaevYG,\KhavaevGB}, 
and a brane-probe study can be found in \refs{\JohnsonIC,\JohnsonZE}.

Our purpose here is to find a family of flows that correspond to
branes with an arbitrary, rotationally symmetric distribution of branes
on this Coulomb branch.  As usual, the vevs of the scalar fields
correspond to directions perpendicular to the branes in the
supergravity solution, and we will use polar coordinates $(u,\phi)$ to
describe the $\Phi_3$-direction, and
$(v,\varphi_1,\varphi_2,\varphi_3)$ to describe the
$(\Phi_1,\Phi_2)$-direction.  The $\varphi_j$ may be thought of as
Euler angles on the $S^3$'s at constant $v$ in the
$(\Phi_1,\Phi_2)$-direction.  The mass deformation in \gaugesuperpot\
preserves an $SU(2) \times U(1) \times U(1)$ subgroup of the original
$SU(4)$ $\cR$-symmetry.  The $SU(2) \times U(1) = U(2)$ acts on
$(\Phi_1,\Phi_2)$ as a doublet, while the last $U(1)$ is a
$\phi$-rotation.  In the finite $N$ field theory this $U(1)$ is
anomalous, but in large $N$ this is restored and it is thus a symmetry
of the supergravity solution.  The brane moduli space is at $u=0$ and
is spanned by $(v,\varphi_j)$, and the solution we seek has the branes
spread out with a density that is an arbitrary
function, $\rho(v)$, of the radial variable on the moduli space.  This
choice keeps the problem relatively simple in that it preserves all
the symmetries.

\subsec{The supergravity background}

The supersymmetry variations for the gravitino, $\psi_M$, and the
spin-${1 \over 2}$ field, $\lambda$, in IIB supergravity read
\SchwarzQR:
\eqn\susytrpsi{
\delta\psi_M\eql D_M\ep+{i\over 480}F_{PQRST}\,\ga^{PQRST}\ga_M\,\ep
+{1\over 96}\left(\ga_M{}^{PQR}-9\,\delta_M{}^P\ga^{QR}\right)\,G_{PQR}
\,\ep^*\,,
}
and
\eqn\susytrla{
\delta\lambda\eql i\,P_M\,\ga^M\epsilon^*-{i\over 24} \,G_{MNP}\,
\ga^{MNP}\,\ep\,,
}
where $\ep$ is a complex chiral spinor satisfying%
\foot{We use the same notation and $\gamma$-matrix conventions as in
\SchwarzQR, except that we label the indices from 1 to 10. Also, see appendix A of \PilchJG.}
\eqn\chirsp{
\ga^{11}\ep\eql-\ep\,.
} 

We now take the metric to have the form:
\eqn\metrans{
ds^2\eql H_1^2 (dx_\mu)^2-H_5^2\,dv^2-H_6^2\,(du^2+ 
u^2\, d\phi^2) -H_7^2\,(\si_1^2+\si_2^2) -(H_{03}\,\si_3+H_{00}\,d\phi)^2\,,
}
and we use the frames:
\eqn\frames{\eqalign{e^a ~=~ & H_1 \,  dx^a \,, \ \ \  a=1, \dots, 4 \,,  \quad  \quad 
e^5 ~=~ H_5 \, dv \,, \quad e^6 ~=~ H_6 \, du \,, \cr
e^7 ~=~ & H_7 \, \sigma_1\,, \quad e^8 ~=~ H_7 \, \sigma_3\,,  \quad
e^{9}  ~=~  -u\, H_6 \,d\phi \,,  \quad e^{10} ~=~  H_{03}\,\si_3+H_{00}\,d\phi\,,}}
where the $H_I=H_I(u,v)$ are functions of both $u$ and $v$,
and the $\sigma_i$ are the left-invariant one-forms parametrized
by the Euler angles $\varphi_i$, $i=1,2,3$, and  normalized so that $d
\si_1 = \si_2 \wedge \si_3$.  

This Ansatz is based upon the form of the holographic field theory
outlined above and the ten-dimensional lift of the LS-flow in \PilchFU. The primary difference
between the result in \refs{\PilchFU} and the Ansatz is that we have introduced the more natural coordinates, $(u,v)$.  Indeed, the solution of
\PilchFU\ exactly fits our Ansatz if one changes variable according to:\foot{ 
As with the $\cN=2$ solutions of \PilchJG, one can show that this change of coordinates 
can be associated with a Lorentz rotation which brings the Killing spinors 
of unbroken supersymmetries into a canonical form.}
\eqn\uvrtheta{ 
u\eql e^{ {3 \over 2} A }\,\sqrt{\sinh\chi}\,\sin\th\,,\qquad 
v\eql\ e^{ A }\,\rh\,\cos\th
 \,.}

One of the important features of the new variables is the simple form
of the metric in the $(u,\phi)$ direction.  Indeed, the metric has a
natural almost-complex structure on the internal space:
\eqn\acstruct{J~\equiv~   - e^6 \wedge e^9 ~+~   e^7 \wedge e^8 ~+~  
e^5 \wedge e^{10} \,.}
Note that there are some arbitrary choices of sign that can be made in
each term.  The choices that we have made here will correlate with
helicity projections that define the supersymmetry.

Since the dilaton and axion backgrounds were trivial in \PilchFU, we
will will also seek such backgrounds here.

Following the observations of \CorradoNV, we make an Ansatz for the two-form 
potential  in which all the indices are holomorphic with respect to \acstruct:
\eqn\Atwopot{
A_{(2)}\eql i\,e^{-i\phi}\left[\,a_1\,(e^6+ie^9)+a_2\,(e^5-ie^{10})\,\right]\wedge
(e^7-ie^8)\,
}
for some functions, $a_1(u,v)$ and $a_2(u,v)$.  
The factor of $(e^7-ie^8)$ in \Atwopot\ is required by 
the action of the $U(1)$ symmetries, but beyond this one could
make a more general Ansatz for $A_{(2)}$.  In principle one
should be able to fix this using the form of the Killing spinors
defined below, combined with the variations \susytrpsi\ and
\susytrla.  In practice, solving such a system is hard, and so
we have made the holomorphic Ansatz above based upon the 
observations of \CorradoNV.

Because the dilaton
and axion are trivial, the three-form field strength, $G_{(3)}$, is simply
\eqn\thegth{ G_{(3)}\eql dA_{(2)}\,. }

The foregoing Ansatz for the three-index tensor in  \Atwopot\ and \thegth\  is  
more restrictive  than the one for the $\cN=2$ supersymmetric solutions in \PilchJG, 
where the only requirement was that certain components, $G_{MNP}$, 
of the three-index field strength vanished.  Here we start with an Ansatz for the potential, 
and the basic reason is the smaller amount of supersymmetry ($\cN=1$) in the present 
problem.   If one makes a very general Ansatz, then the  system of Killing equations  and  
Bianchi identities that would result  here would considerably more  difficult to analyze
than that encountered in \PilchEJ.

Finally, we define:
\eqn\fourindpot{
C_{(4)}\eql w\,dx^0\wedge dx^1\wedge dx^2\wedge dx^3\,, }
for some function, $w(u,v)$, and then take the five-form field strength to be:
\eqn\fiveform{ F_{(5)}\eql dC_{(4)}+*dC_{(4)}\,. }

\subsec{The supersymmetries}

Having made the Ansatz for the metric and the tensor fields, 
we now restrict the form of the Killing spinors through sets of projectors. 
First, there is the helicity  condition \chirsp\ on the spinors of the IIB theory:
\eqn\chiral{
\Pi_{11}\,\epsilon\eql\epsilon\,, \qquad {\rm where} \qquad  
\Pi_{11}\eql {1\over 2}\left(\cI-\ga^{11}\right)\,.
}

Next, we follow the philosophy outlined in \GowdigereJF, and see how the
supersymmetry must be defined on the moduli space of the brane probes
and then assume that whatever projection conditions are needed on that
space will lift, without modification, to the full space.  To reduce
to one-quarter supersymmetry on the four-dimensional moduli space one
must reduce the four-component spinors to a single helicity component.
That is, one needs to fix the helicity of $\gamma^{78}$ and
$\gamma^{5\,10}$. These pairs of $\gamma$-matrices are the natural
ones given by the almost complex structure of \acstruct.  We therefore
introduce the projectors:
\eqn\projectone{
\Pi_{78}\eql {1\over 2}\left(\cI-i\,\ga^{7}\ga^8\right)\,,\qquad
\Pi_{5\,10}\eql {1\over 2}\left(\cI-i\,\ga^{5}\ga^{10}\right)\,,
}
and impose the further conditions:
\eqn\projcond{
\Pi_{78}\,\epsilon \eql\epsilon \,,\qquad
\Pi_{5\,10}\,\epsilon \eql\epsilon \,.
}
There are choices of sign to be made in the definitions of
\projectone.  As we will see, the choices here are fixed by the
choices of signs in \acstruct, \Atwopot\ and \frames.

The final projector is a dielectric deformation of the standard projector 
for the $D3$-branes:
\eqn\nontrivproj{
\Pi_{1234} \eql
{1\over 2}\left[\cI+i\,\ga^1\ga^2\ga^3\ga^4(\cos\be-\,e^{-i\ph}\sin\be
\,\ga^7\ga^{10}\,*\,)\right]\,,
}
where $\beta=\beta(u,v)$ is a function to be determined.   We then impose the 
condition 
\eqn\lastcond{
\Pi_{1234} \,\epsilon\eql\epsilon\,.}

We will describe below how we arrived at this projector, however its
form is essentially fixed by the physics and mathematics of the
problem.  First, because we are dielectrically polarizing the
$D3$-branes into $D5$-branes and $N\!S5$-branes, the deformation term 
must be a product
of six $\ga$-matrices (for the five-branes) containing
$\ga^1\ga^2\ga^3\ga^4$ (for the $D3$-branes).  Thus we need to find
the two extra $\ga$-matrices.  The result must commute with
$\Pi_{78}$ and $\Pi_{5\,10}$, and be a ``true deformation.''  For
example, $\ga^1\ga^2\ga^3\ga^4 \ga^7\ga^8$ will not suffice because
\projcond\ means that  it  is the same as $\ga^1\ga^2\ga^3\ga^4$.  This
leads one to choose one of $\ga^7, \ga^8$ and one of  $\ga^5, \ga^{10}$, and
\projcond\  means that it does not matter which ones we choose.  Having made
the choice, the complex conjugation, $*$, operation is essential for
$\Pi_{1234}$ to commute with $\Pi_{78}$ and $\Pi_{5\,10}$.  Finally,
the factor of $e^{-i\ph}$ is required to correct the $\phi$-dependence
after complex conjugation.   Put more physically, $\phi$-rotations
generate the $U(1)$ $\cR$-symmetry, and the $e^{-i\ph}$ term is
essential for the projector to preserve $\cR$-symmetry.
 
It is worth noting that the projection conditions \projcond\ and \nontrivproj\
are natural generalizations of the corresponding conditions 
used to define the Killing spinors on a Calabi-Yau manifold.

Having defined the space of supersymmetries, we need to fix
their dependence on the various coordinates.   The angular dependence can be
fixed using the Lie derivative on spinors \refs{\TownsendCI,\FigueroaVA}:
\eqn\LieDeriv{L_{K}\, \epsilon ~\equiv~ K^M \, \nabla_M \, \epsilon
~+~ \coeff{1}{4}\, \nabla_{[M} K_{N]} \, \gamma^{MN} \, \epsilon \,.}
The fact that the spinors are singlets of the $SU(2)$ symmetry means
that $L_{K}\, \epsilon = 0$ for Killing vectors, $K^M_{(j)}$, in these
directions.  If one chooses the $\si_j$ with the appropriate
handedness, the connection term and the $\nabla_{[M} K_{N]} $ term
cancel, and one is left with $\partial_{\varphi_j}
\epsilon=0$, $j=1,2,3$.  A similar cancellation takes place for the Killing
vector in the $\phi$-direction, except that this is the residual
$\cR$-symmetry that acts on the supersymmetry, and $\epsilon$ has
charge $+{1 \over 2}$.  The tensor gauge field \Atwopot\ has
$\cR$-charge $+1$, and therefore one has:
\eqn\epsphi{\partial_\phi  \, \epsilon  ~=~  \gamma^6\gamma^9   \,
\epsilon  ~-~ \coeff{i}{2}\, \epsilon \,.}
The $\phi$-phase dependence in \nontrivproj\ can also be fixed by 
requiring that the projector commute with this Lie derivative operator.

Finally, the dependence on $(u,v)$ can be fixed by using the fact that
$K^M ~\equiv~ \bar\epsilon \gamma^M \epsilon$ is always a Killing
vector.  Using spinors that satisfy the projection conditions, one
finds that $K^M$ is either zero, or it is parallel to the brane, in
which case it must be a constant vector.  This fixes the normalization
of $\epsilon$ in terms of $H_1$.  The end-result is that we have
completely determined the form of supersymmetries up to an arbitrary function, $\beta$.

We conclude by observing that the foregoing can be re-written rather
more directly by introducing the ``rotation'' matrix:
\eqn\defofbe{
\cO^*(\be)\eql \cos({\be\over 2})+\sin({\be\over 2})\,\ga^7\ga^{10}\,*\,.
}
The Killing spinor is then given explicitly by:
\eqn\killspinor{
\epsilon\eql H_1^{1/2}\,e^{-i\phi/2}\,\cO^*(\beta)\,e^{i\phi}\epsilon_0\,,
}
where $\epsilon_0$ is a {\it constant} spinor satisfying:
\eqn\simpprojcond{
\Pi_{78}\,\epsilon_0 \eql\epsilon_0 \,,\qquad
\Pi_{5\,10}\,\epsilon_0 \eql\epsilon_0 \,, \qquad
\Pi^{(0)}_{1234} \,\epsilon_0 \eql\epsilon_0 \,,
}
and where
\eqn\simpprojdefn{
\Pi^{(0)}_{1234} ~\equiv~ {1\over 2}\left[\cI + i\,\ga^1\ga^2\ga^3\ga^4 \right] \,.}
Conjugating $\Pi^{(0)}_{1234}$ by $e^{-i\ph/2}\cO^*(\be)$ results in the deformed
projector \nontrivproj,
\eqn\conjproj{\Pi_{1234}\eql e^{-i\ph/2}\,
\cO^*(\be)\,\Pi^{(0)}_{1234}\,\cO^*(\be)^{-1}\,e^{i\ph/2}\,.
}  Thus the whole family of solutions considered
here may be thought of as a duality rotation of a standard, $\cN=1$
supersymmetric brane compactification whose supersymmetries are
defined by \simpprojcond.

\newsec{Solving the Ansatz}

We now use our Ansatz in \susytrpsi\ and \susytrla\ and solve for all the 
undetermined functions.  

\subsec{The step-by-step process}

The easiest one to solve is \susytrla\ which, because of the trivial
dilaton/axion background, collapses to $G_{MNP}\,\gamma^{MNP} \epsilon
= 0$.  This leads immediately to the projection condition
$\Pi_{78}\epsilon =\epsilon$.  Next, we observe that in the combination
\eqn\nicecomb{
2\ga^1\delta\psi_1+\ga^7\delta\psi_7+\ga^8\delta\psi_8\eql 0}
all terms with the antisymmetric tensors cancel and the entire
contribution on the left-hand-side comes from the spin connection.  It
is this that leads us to the second projection condition,
$\Pi_{5\,10}\epsilon =\epsilon$.  We thus see how the choices of signs
in the Ansatz result in the sign choices for the projectors \projectone.  From
this we also obtain the conditions:
\eqn\specvarsol{
H_{03}\eql  {v\over 2\,H_1^2\,H_5}\,, \qquad \partial_u(H_1H_7)\eql 0\,.}
Motivated by this we set:
\eqn\hseven{
H_7=\coeff{1}{2}\, v \, H_1^{-1} \,.
}
We are free to choose this solution because we have not yet fixed the
coordinate freedom in $u$ and $v$.  Specifically, we can redefine $v
\to 2 H_1H_7$, and then redefine $u$ so as to remove any $du\,dv$
cross-terms in the metric.  The metric Ansatz is now reduced to:
\eqn\metranz{\eqalign{
ds^2\eql H_1^2 (dx_\mu)^2-H_5^2\,dv^2&-H_6^2\,(du^2+u^2d\phi^2)\cr& -
{v^2\over 4H_1^2}\,(\si_1^2+\si_2^2)-{v^2\over 4 H_1^4H_5^2}\,(\si_3+2H_0\, d\ph)^2\,,
\cr}
}
 where $H_0 \equiv v^{-1}  H_{00}\,H_1^2\,H_5 $.

The remaining supersymmetry variations yield an entangled   
system of first order differential equations. However, by taking suitable linear 
combinations of those equations it is quite straightforward to  determine $H_1$ 
and $w$ in terms of the other functions in the Ansatz:
\eqn\theoneh{
H_1^2\eql {1\over 2\,v \,H_5\,H_6}{\partial\over\partial v}\left({v^2\,H_6\over H_5}\right)\,,\qquad 
w\eql -{1\over 4} \,H_1^4\,\cos\be\,.
}

At this point the algebra gets tougher. In addition,  one needs to 
consider separately the two cases depending on whether $\beta$ is zero or not. 
In the appendix we have summarized the remaining equations that one must still 
solve at this point.

If  $\beta=0$ then the solution is degenerate in the sense  that the supersymmetry 
variations alone do not determine all the functions in the Ansatz. That is,
 the $\cN=1$ supersymmetry does not determine a solution without a 
complete analysis of the field equations. This also happens in the more
familiar, standard harmonic solutions with unbroken supersymmetry.  We 
refer the reader to the appendix for  some additional discussion of the
solution with $\beta=0$  and in the following  assume that  $\beta\not=0$.

For $\beta\not=0$, the main conclusion of a detailed 
analysis of the equations in the appendix is that using the supersymmetry 
variations one can determine explicitly all functions in the Ansatz in terms of 
just two of them, $H_5$ and $H_6$. In particular, in addition to \theoneh, we have
\eqn\sinbeta{
\sin\beta\eql {2\, u\over H_1^3 \,H_6}\,.}
Then, using \theoneh, we can integrate (A.9) to obtain 
\eqn\thehostuff{
H_0 \eql -\ccoeff{1}{2} \,u\, {\partial\over \partial u}\,\log\left({H_6 \over H_5}\right)\,. }
Finally, the two-index potential is given by 
\eqn\thehone{
a_1\eql {v\,H_1\,H_0\over u^2\,H_5}\,,\quad a_2\eql-\tan(\coeff{1}{2}
\beta)\,.}

\subsec{Building the solution}

\def\cH{{\cal S}}

To solve for $H_5$ and $H_6$ one must disentangle the remaining  
first order system 
of differential equations.  It is convenient to define:
\eqn\Psidefn{
\Psi\eql \log\left(v^2\,{H_6\over H_5}\right)\,.
}
This function must satisfy the ``master equation:''
\eqn\mastereqn{
u^3 {\partial\over\partial u}\left({1\over u^3}\,
{\partial\over\partial u}\,\Psi\right)+
{1\over 2 v}\,
 {\partial\over\partial v}\left({1\over v^3}\,
 {\partial\over\partial v}\, e^{2\Psi}\right)\eql 0\,.
}

Associated with $\Psi$ there is a conjugate function, $\cH$, defined by:
\eqn\fstorsys{
{\partial\cH\over\partial u}\eql -{1\over 2 \,v^3\,u^3}\,{\partial e^{2\Psi}\over\partial v}\,,
\qquad {\partial\cH\over\partial v}\eql {v\over u^3}\,{\partial \Psi\over\partial u}\,.}
The ``master equation'' is the integrability condition for $\cH$.  If
one solves \mastereqn\ and integrates the solution to obtain $\cH$,
one can then determine all other functions as follows.  First one
determines $\beta$ from:
\eqn\thetanbeta{
\tan^2(\coeff{1}{2} \,\beta )\eql -\big (1+\coeff{1}{2}\,u\, \partial_u \log\cH)\eql
{e^{2\Psi}\,\partial_v\Psi\over 2\,u^2\,v^3\,\cH} 
\eql {\partial_v e^{2\Psi}\over 4\,u^2\,v^3\,\cH }\,.
}
Then one has:
\eqn\fixHfns{{H_6\over H_5} ~=~ {1 \over v^2}\, e^\Psi \,, \qquad
H_1^3\,H_6 \eql {2\,u\over  \sin\beta }\,, \qquad
H_1^2\,H_5^2 \eql \coeff{1}{2}\, v\,{\partial\Psi\over\partial v} \,,
}
from which one can algebraically determine $H_1$, $H_5$ and $H_6$.  In
particular, one can  see that 
\eqn\thehone{
H_1^4 = {u^2 \over v^3 \, \sin^2
\beta}\, \partial_v (e^{2\,\Psi}) \,,
}
 and then one can easily read off
$H_5$ and $H_6$. The remaining functions are then given by:
\eqn\fixhzero{
H_0  \eql  -\coeff{1}{2} \, u\,{\partial\Psi\over\partial u} \,,}
\eqn\fixas{
a_1 \eql {v\,H_1\,H_0\over u^2\,H_5}\,,\qquad a_2 \eql-\tan(\coeff{1}{2}
\beta)\,,}
\eqn\fixw{
w  \eql -\coeff{1}{4} \,H_1^4\,\cos\beta\,. 
}
Thus, once one solves \mastereqn, one has solved the entire Ansatz.
 
We have also verified that given these equations, the Ansatz satisfies 
 all the Bianchi identities and the equations of motion of the IIB theory.
 
 \subsec{A comment on the perturbation expansion}
 
Finally, we note that for solutions that 
 are asymptotic to $AdS_5 \times  S^5$, or asymptotic to any
 Coulombic brane distribution, one wants $H_5 \to H_6$, or
 $\Psi \to \log(v^2)$.  Define $\tilde \Psi \equiv \Psi -  \log(v^2)$
 and observe that $\tilde \Psi \to 0$ at infinity and that $\tilde \Psi$ satisfies 
 the master equation if and  only if $\Psi$ satisfies the same equation. 
 This means that we can make a simple asymptotic perturbation expansion
 for $\tilde \Psi$:
 \eqn\pertexp{\tilde \Psi(u,v)  ~=~ \sum_{n=1}^\infty ~ \xi^n ~\chi_n(u,v) \,,}
where $\xi$ is a small parameter and the $\chi_n$ are to be determined.
The function $\chi_1$ satisfies the linearized master equation:
\eqn\linmaster{\cL\,\chi_1  ~\equiv~ 
u^3 {\partial\over\partial u}\left({1\over u^3}\,
{\partial\over\partial u}\,\chi_1 \right)+
{1\over v}\,
 {\partial\over\partial v}\left({1\over v^3}\,
 {\partial\over\partial v}\, \chi_1 \right)\eql 0\,,
}
while the $\chi_n$ must satisfy a {\it linear} equation of the form:
\eqn\sources{\cL\,\chi_n  ~=~  \cP(\chi_1,\dots,  \chi_{n-1})\,,
}
for some polynomial, $\cP$.  There is, of course, the ambiguity of
adding in more of the homogeneous solution at each step, but such
ambiguities are easily resolved in terms of a re-defintion of $\chi_1$.
The function $\chi_1$ is thus the ``seed function'' for the complete 
solution.  Since it satisfies the linear PDE \linmaster, we may choose
it to have a source defined by a function, $\rho(v)$, in the plane $u=0$.
While this function may not ultimately be true brane distribution of
the non-linear problem, our argument shows that while we have a non-linear
problem, there is a good perturbation theory, and that there is indeed
a family of solutions determined by the choice of an arbitrary function $\rho(v)$.
Moreover, this function will determine the multipole
expansion of the brane distribution as seen from infinity.

\newsec{Some special solutions}

\subsec{The FGPW-flow}

The flow solution obtained in \FreedmanGP\ was based upon five-dimensional supergravity
and involved the metric:
\eqn\RGFmetric{
ds^2 = e^{2 A(r)} \eta_{\mu\nu} dx^\mu dx^\nu - dr^2 \ .}
The equations of motion were:\foot{In the following we set $L=1$.}
\eqn\RGFeqns{{d \varphi_j \over d r} ~=~ {1 \over L}~{\del W \over \del \varphi_j} \,, 
\qquad {d A \over d r} ~=~ - {2 \over 3\,L}~W \ ,}
with
\eqn\Wreduced{W~\equiv~ {1 \over 4 \rho^2}~ \Big[\cosh(2 \chi)~
( \rho^{6}~-~ 2)~ - ( 3\rho^{6} ~+~ 2 ) \Big] \,,}
where $\chi=\varphi_1,  \rho=exp( {1 \over \sqrt{6}} \varphi_2)$.  

The lift of this solution to ten dimensions was presented in \PilchFU\ and further simplified in \CorradoNV. As remarked above, the
ten-dimensional solution is obtained from our Ansatz by taking:
\eqn\therttouv{
u(r,\theta)\eql e^{ {3 \over 2} A }\,\sqrt{\sinh\chi}\,\sin\th\,,
\qquad v(r,\theta)\eql\ e^{ A }\,\rh\,\cos\th
 \,.
}
For completeness we also note that one has:
\eqn\theLSPsiandH{
\Psi(u,v)\eql\log\left({e^{ {3 \over 2} A }\,\cosh\chi\,\cos^2\th\over
\sqrt{\sinh\chi}}\right)\,,\qquad
\cH(u,v)\eql{e^{ -4\, A } \over  \rh^4\,\sinh^2\chi\,\sin^2\theta}\,.
}
One can then solve \therttouv\ and \theLSPsiandH\ to obtain
\eqn\LSidentit{
 \cosh\chi  \eql {e^\Psi\over u\,v^2\,\sqrt\cH}\,\,,\qquad 
{\cos^4\th\over\sin^2\th}\eql {e^{2\Psi}\over u^2}-v^4\,\cH\,,
}
but there is no similarly simple formula to express $\rho$ as a function of $u$ and $v$.

\subsec{The KPW fixed point}

The general flows considered here should correspond to the Coulomb branch
of the $\cN=1$ supersymmetric fixed point theory described in \LeighEP.  This means
that the supergravity solution will generically be singular in its core, and
the singularity will merely reflect the appropriate continuum distribution of branes.
There is, however, one non-singular background, discovered in \refs{\KhavaevFB,\PilchEJ} 
that represents the conformal IR fixed point field  theory \refs{\KarchPV,\FreedmanGP}, and in which
the space-time has an  $AdS_5$ factor.   In terms of the parameterization 
above, this point has: 
\eqn\thepoint{
\cosh\chi(r)\eql {2\over\sqrt3}\,,\qquad \rho(r)\eql 2^{1/6}\,,\qquad
A(r)\eql {2\,2^{2/3}\over 3}\,r\,,
}
which gives:
\eqn\cruv{
u\eql {1\over 3^{1/4}}\,e^{2^{2/3}\,r}\,\sin\th\,,\qquad
v\eql 2^{1/6}\,e^{(2\,2^{2/3}/3)\,r}\,\cos\th\,.
}
The master function is
\eqn\mstfnct{
\Psi(u,v)\eql \coeff{1}{2}\,\log\Big(v^3\,F\Big({9\over\sqrt2}\,{u^2\over v^3}\Big)\Big)\,,
}
where
\eqn\subcrmf{
F(x)\eql {\sqrt2\over x}\, \left( {\left( x + {\sqrt{x^2-1}} \,\right) }^{2/3}+
 {\left( x-{\sqrt{x^2-1}} \,\right) }^{2/3}-1\right) \,.
}
One then finds that:
\eqn\fstcrfn{
\cH(u,v)\eql {27\over 2\,v^4}\,H\Big({9\over\sqrt2}\,{u^2\over v^3}\Big)\,,
}
where
\eqn\fnctcrh{
H(x) ~\equiv~ {\left( x + {\sqrt{x^2-1}}\, \right)^{1/3}\over
x\,\left(1+\left( x + {\sqrt{x^2-1}}\, \right)^{1/3}\right)}\,.
}

\subsec{Other solutions}

Motivated by the form of  \mstfnct, one can seek solutions of the form:
\eqn\otheraa{
e^{2 \Psi(u,v)}\eql v^a u^b\,\cF(v^c u^d)\,.
}
Then the function $\cF(x)$ satisifes an ordinary differential equation  provided
\eqn\otherab{
a\eql 6+{c\,(2+b)\over d}\,,
}
and the coefficients in the eqs have only integer powers of $x$ when
\eqn\otherac{
{b+2\over d}\eql n\,,\qquad n\in\ZZ\,.
}
The critical point solution has
\eqn\otherad{
a\eql 3\,,\qquad b\eql 0\,,\qquad c\eql -3\,,\qquad d\eql 2\,,\qquad n\eql 1\,.
}
Another solution in this spirit is obtained by setting 
\eqn\otherae{
a\eql 2\,,\qquad b\eql c\eql -2\,,\qquad d\rightarrow 0\,,\qquad n\eql 2\,,
}
and is simply
\eqn\simpsol{
e^{2\Psi(u,v)}\eql -{2\over 3}\, {v^6\over u^2}\,.
}
Here the differential equation is linear:
\eqn\otheraf{
x^4\cF''-3x^3\cF'+8\eql 0\,. 
}
For $d=0$ we have to divide out some singular terms  to arrive at the
differential equation \otheraf,
and it turns out that only the solution \simpsol\  gives rise 
to a solution of the master equation.

Finally, there are obvious solutions  where we set each term in the 
master equation to zero.  This yields solutions in which:
\eqn\specsolpsi{
\Psi(u,v)\eql {1\over 2}\log(c_1+v^4)+c_2 u^4+c_3\,.
}
The corresponding solution for $\cH$ is then:
\eqn\specsolhh{
\cH(u,v)\eql {1\over 2}c_2v^2-e^{2c_3}\sqrt{{\pi c_2\over 2}}\,{\rm Erfi}\left(
\sqrt{c_2\over 2}\,u^2\right)+
{\hbox{$\displaystyle {e^{2c_3+c_2u^4/2}}$}\over u^2}+c_4\,.
}
where $c_1$, $c_2$, $c_3$ and $c_4$ are integration constants. 

\newsec{Final comments}

We have shown that the algebraic Killing spinor techniques proposed in
\refs{\PilchJG,\GowdigereJF} can be successfully applied to problems with fewer
supersymmetries.  The ideas adapt very directly to the new class of
problems considered here, and yield infinite families of solutions
that once again generalize the harmonic Ansatz.  We therefore believe
that these techniques will find broader applications within string
theory, and as we indicated in the introduction, research on these
issues is continuing.

On a more physical level, we have once again seen that interesting
supersymmetric flow solutions involve a dielectric polarization of
$D3$-branes into five-branes. This polarization does not break the
supersymmetry by itself: The supersymmetry projector undergoes a
duality rotation in which the original $D3$-branes directions are all
parallel to the emergent distribution of five-branes.  Indeed, this
represents a unifying thread between this paper and our earlier work:
In more standard compactifications the supersymmetries are defined by
simple geometric projection conditions, and we are essentially
replacing the canonical projector associated with the branes by some
dielectric deformation, while leaving all the other projectors
untouched.  Since the solutions we are studying are holographic duals
of interesting flows in supersymmetric gauge theories, we believe that
these dielectric deformations of branes will play a significant role
in understanding supersymmetric compactifications and supersymmetry
breaking within string theory.

\bigskip
\leftline{\bf Acknowledgements}

This work was supported in part by funds
provided by the U.S.\ Department of Energy under grant number 
DE-FG03-84ER-40168. 
In addition,  K.P. has been supported in part by the grant \# SAB2002-0001 from the 
Ministerio de Educaci\'on, Cultura y Deporte of Spain.  K.P. would like to thank Jos\'e 
Latorre and all the members of the Departament ECM  for their hospitality during his 
sabbatical at the University of Barcelona.
N.W. would like to thank
Iosif Bena for helpful discussions.
\vfill\eject

\appendix{A}{Supersymmetry constraints}

In this appendix we summarize the  supersymmetry equations which remain when the 
partial solutions \specvarsol, \hseven\ and \theoneh\ are used in  the Killing spinor 
equations \susytrpsi\ for the $\cN=1$ supersymmetry. 

We observe that  Ansatz for the two-form potential \Atwopot\ implies, in particular, 
that the corresponding field strength is of the form
\eqn\asymmansatz{\eqalign{
G_{(3)}\eql \big(&i\,g_{56}\,e^5\wedge e^6+i \,g_{90}\,e^9\wedge e^{10}\cr &
+g_{59}\,e^5\wedge e^9+g_{50}\,e^5\wedge e^{10}+g_{69}\,e^6\wedge e^9+g_{60}\,
e^6\wedge e^{10}\big)\wedge (e^7-i\,e^8)\,,\cr}
}
where all functions $g_{\omega}(u,v)$ are real. Since the supersymmetry 
variations constitute a system of linear equations for the components of the three-index 
field strength, it is convenient to first determine $g_\om$'s in terms of the other 
functions in the Ansatz. It is at this point that  the $\beta\not=0$ case differs
from that of $\beta=0$.

\subsec{Solving with  $\beta\not=0$}

One finds that all the functions $g_\om$ are determined by the supersymmetry 
variations through the following system of equations:
\eqn\asymeqs{\eqalign{
g_{56}-g_{60} & \eql {1\over H_6}\,{\partial\beta\over\partial u}\,,\qquad 
g_{59}+g_{90}\eql -{1\over H_1^4H_6}\,{\partial\over\partial u}(H_1^4\sin\beta)\,,
\cr
g_{50} &\eql  -{1\over H_5}\,{\partial\beta\over \partial v}\,,\qquad
g_{69}\eql {1\over H_1^4H_5}\,{\partial\over\partial v}(H_1^4\sin\beta)\,,\cr}
}
and
\eqn\gfivesix{
3 g_{56}+g_{60}\eql {H_1^2\over 2\,u^2\,v\,H_5^2}\left(
{2v^2\over H_1}\,{\partial\over\partial v}( {H_0} )
-{uv\over H_1^3H_5^2}{\partial\over\partial u}\,\big( H_1^4H_5^4\cos\be \big)
\right)\,,
}
\eqn\gninezero{\eqalign{
g_{90}-3 g_{59}\eql{H_1^2\over 2\,u^2\,v\,H_5^2}\,\Big(
&{2 v^2\cos\be\over H_1}{\partial\over\partial v} ( {H_0} )
\cr &\quad -{uv\cos\be\over H_1^3H_5^2}{\partial\over\partial u}\big( H_1^4 H_5^4\cos\beta
\big)-{8 u^3 v \over H_1^5H_6^2}{\partial\over\partial u}\big(H_5^2\big)\Big)\,.
}}
Here $H_1$ is given by \theoneh\ and, in addition, one finds that 
\eqn\thebeta{
\sin\beta\eql {2\, u\over H_1^3 \,H_6}\,.
}
Then the following two equations 
\eqn\delpsisix{
{12\,  v\over H_1^4H_6}\,{\partial\over\partial u}( H_0)+
{u^8\,H_1\over 2 H_6^6}{\partial\over\partial v}\big({H_6^8\sin(2\be)\over u^8}\big)
-5\,H_1H_6^2{\partial\beta\over\partial v}\eql0\,,
}
and
\eqn\lasteqs{
{v\over 2 \, H_1^2H_5^2}\,{\partial\over\partial v}(H_0)-
{u^8\,H_1^3\over 48\,H_6^7}{\partial\over\partial u}\big({H_6^8\sin(2\be)\over u^8}\big)
+2\,H_0+{5\over 24}\,H_1^3H_6\,{\partial\beta\over\partial u}\eql 1\,,
}
exhaust the remaining sypersymmetry constraints.

In deriving \asymeqs-\lasteqs\ we have only used the general form of $G_{(3)}$ 
in \asymmansatz. An additional  restriction on the functions $g_\om$, that follows 
from the holomorphic Ansatz for the two-form potential  \Atwopot, is
\eqn\curident{
g_{60}+g_{59}+g_{56}-g_{90}\eql 0\,,
}
and it brings a dramatic simplification of the problem. Indeed, if we
substitute \asymeqs-\gninezero\ in \curident\ and then use 
\delpsisix\ and \lasteqs, we obtain
a remarkably simple result
\eqn\simres{
{\partial\over\partial v}H_0\eql -{u\over v}\,{\partial\over\partial u}
(H_1H_5)^2\,,
}
which is crucial for solving the Ansatz in section 3.

\subsec{Solving with $\beta=0$}

For $\beta=0$, several equations become dependent and the supersymmetry 
variations alone do not determine all fields in the Ansatz. In particular, the functions 
$g_\om$  satisfy
\eqn\thebzergs{
g_{50}\eql g_{69}\eql 0\,,\quad g_{56}\eql -g_{59}\eql g_{60}\eql g_{90}\,,
}
and thus \curident\ does not provide any additional constraint. 
The remaining supersymmetry variations consist of three 
equations which can be written as 
\eqn\rembezer{
{\partial\over \partial  u} H_0\eql - {u\over v}\,{\partial\over\partial v} (H_1H_6)^2\,,\qquad
{\partial\over \partial  v} H_0\eql  {u\over v}\,{\partial\over\partial v} (H_1H_5)^2\,,\qquad
}
together with\foot{The reader might be puzzled about the sign difference between 
\fixhzero\ and (A.12). However, there is no contradiction here, as the initial set of equations
 that lead to \fixhzero\ and (A.12) are different for $\beta\not=0$ and $\beta=0$, respectively.} 
\eqn\thehzero{
H_0\eql {1\over 2}\,u\,{\partial \Phi\over \partial u}\,,
}
where $\Phi$ is the ``master function''
\eqn\Phidefn{
\Phi\eql \log\left(v^2\,{H_6\over H_5}\right)\,.
}
A consistency between \rembezer\ and \thehzero\ using \theoneh\ requires that $\Phi$ 
satisfies the master equation
\eqn\newmaster{
{1\over u}\,{\partial\over\partial u}\left(u\,{\partial\over\partial u}\,\Phi\right)+
{1\over 2v}\,{\partial\over\partial v}\left( {1\over v^3} {\partial\over \partial v}e^{2\Phi} 
\right)\eql 0\,,
}
which is different from \mastereqn, which applies for $\beta\not=0\,$!

To summarize, we find that, unlike in the $\beta\not=0$ case,  the supersymmetry 
variations leave one function in the metric, e.g., $H\equiv H_5^4$, and the two-form 
potential independent  of the master function $\Phi$.

\bigbreak
\noindent
{\it Example:} $AdS_5\times S^5$

An obviously important example of the $\beta=0$ case is the $AdS_5\times S^5$ 
compactification. It is easy to check that this solution corresponds to 
\eqn\theuv{ 
\Phi(u,v)\eql \log(v^2)\,,\qquad H(u,v)\eql {1\over (u^2+v^2)^2}\,,\qquad g_\omega(u,v)\eql 0\,,
}
where $H$ satisfies  the harmonic equation
\eqn\uvhar{
{1\over v^3}{\partial\over \partial v}\left(v^3{\partial\over \partial v}\,H\right)+
{1\over u}{\partial\over \partial u}\left(u{\partial\over \partial u}\,H\right)\eql 0\,,
}
which follows from the Bianchi identity for $F_{(5)}$ or, equivalently, from the 
Einstein and/or the Maxwell
field equations.

\listrefs

\vfill\eject\end